\documentclass[12pt]{iopart}

\usepackage{iopams}
\usepackage{braket}
\usepackage[utf8]{inputenc}
\usepackage{color}
\usepackage{hyperref}

\newcommand{\dd}{\mathrm{d}}
\newcommand{\ee}{\mathrm{e}}

\newcommand{\vect}{\boldsymbol}
\usepackage{graphicx}
\usepackage{hyperref}

\begin{document}
\title{Statistical topology of the streamlines of a two-dimensional flow}
\author{Mason Kamb$^1$, Janie Byrum$^1$, Greg Huber$^1$,
Guillaume Le~Treut$^1$, Shalin Mehta$^1$, 
Boris Veytsman$^{2,3}$, David Yllanes$^{1,4,\dagger}$}

\address{$^1$ Chan Zuckerberg Biohub, San Francisco, CA 94158, USA}
\address{$^2$ Chan Zuckerberg Initiative, Redwood City, CA 94063, USA} 
\address{$^3$ School of Systems Biology, George Mason University, Fairfax, VA 22030, USA}
\address{$^4$ Instituto de Biocomputaci\'on y F\'{\i}sica de Sistemas Complejos (BIFI), 50018 Zaragoza, Spain}
\address{$^\dagger$ david.yllanes@czbiohub.com}

\date{\today}
\begin{abstract}
  Recent experiments on mucociliary clearance, an important defense
  against airborne pathogens, have raised questions about the topology
  of two-dimensional (2D) flows, such as the proportion of topologically closed and open streamlines.  We introduce a framework for studying ensembles of 2D
  time-invariant flow fields and estimating the probability for a
  particle to leave a finite area (to clear out). We establish two
  upper bounds on this probability by leveraging different insights
  about the distribution of flow velocities on the closed and open
  streamlines. We also deduce an exact power-series expression for the
  trapped area based on the asymptotic dynamics of flow-field
  trajectories and complement our analytical results with numerical
  simulations.
\end{abstract}
\submitto{\jpa}

\section{Introduction}
The study of the properties of random flows is long established in the
field of fluid modeling, particularly in the area of turbulence
\cite{constantin1994geometric,she1991intermittency, KondevHuber2001}. Dating back at
least to Kolmogorov's~1941 studies,
\cite{kolmogorov1991local,kolmogorov1991dissipation}, the use of
statistical ensembles to model characteristics of interest in fluid
flows has achieved widespread success. In the cases where it may not
be possible or desirable to specify detailed structure of a fluid
flow, the statistical methods allow the averaging over the irrelevant
features in order to interrogate quantities of interest. Another
method to abstract away irrelevant details is the use of the topology
of streamlines. In an ideal non-viscous flow, the topology of
streamlines remains invariant in time in Eulerian coordinates. This
means that the properties of the streamlines can provide useful
information about the flow: for example, helping to determine minimal
energy states for a flow subject to topological constraints
\cite{khesin2005topological,moffatt2013topological}. Statistical topological
features of random functions have also been studied from a more abstract
perspective \cite{sarnak2019topologies}, as well as for the purpose of
understanding the features of the Cosmic Microwave Background
\cite{feldbrugge2019stochastic, adler2017modeling}.

 In two dimensions, the topological question is relatively simpler, as there
are two available streamline topologies: `open' and `closed.' Understanding the
topology of two-dimensional flows is essential
for important biological applications. For instance, one key
biophysical mechanism employed in the body's defense to respiratory
pathogens is mucociliary clearance~\cite{fahy2010airway}. In humans
and many other animals, the airway is covered by a thin, quasi-2D
layer of mucus, which acts as a barrier to airborne pathogens. This
mucus layer is actively `cleared' by a layer of ciliated cells. The
beating of these cells generates a quasi-static two-dimensional flow
field in the mucus. This flow is directed \cite{liron1978fluid}, but
exhibits significant spatial structure due to the disordered
arrangement of ciliated cells \cite{rayner1996ciliary,lee2011muco} and
the properties of the beating itself
\cite{guo2014cilia,vilfan2006hydrodynamic}. It has been shown that
an increased disorder in this system is associated with pathological
outcomes in humans~\cite{rayner1996ciliary}. Recent work by
Ramirez-San Juan \emph{et al.}~\cite{RamirezSanJuan2020} has
investigated the characteristics of disordered flows in the airway
system.  These authors show that the proportion of closed streamlines
in the flow negatively impacted the ability to clear viral particles,
which became `trapped' in their orbits rather than being advected out
of the system. Providing a precise characterization of the statistics
of trapping in 2D flow may bring clarity about the magnitude of this
effect. This might provide insight into the pathologies caused by
increasing trapping.

There is a strong connection between the study of random flows and the
study of magnetic fields in disordered materials. The zero-divergence
condition present in both incompressible flows (due mass conservation)
and magnetic fields (due to the absence of magnetic monopoles) means
that the formal techniques used for one problem often also work for
the other. In this context, the problem of identifying the trapped
area in a magnetic field is closely related to certain variations of
the Hall effect
\cite{isichenko1992percolation,trugman1983localization,GORDEEV1994215}. Electrons
moving in a disordered material tend to follow magnetic field lines.
As magnetic bias is provided, more and more streamlines become open,
leading to a non-zero net flow.
Isichenko~\cite{isichenko1992percolation} studied the proportion of
topologically open versus topologically closed streamlines in a biased
magnetic field as a function of the control parameter $\gamma = \mu/\sigma$, where $\mu$ is the magnitude of the mean field and $\sigma^2 = \sigma_x^2 + \sigma_y^2$ is the sum of the componentwise variances. According to \cite{isichenko1992percolation}, the critical
point is $\gamma = 0$, and near this point the fraction $a$ of the
area inside closed streamlines (the `trapped area') scales as
$1 - a \propto \gamma^{4/7}$.  The scaling exponent $4/7$ for this un-trapped area is
related to the critical exponents of percolation theory
\cite{isichenko1992percolation, zhao1993electron,
smirnov2001critical,KondevHuber2001,
KondevHenley1995} through the relation
$c=1/D_\mathrm{h}=\nu/(1+\nu)$, where $D_\mathrm{h}=1+1/\nu$ is the
cluster-hull dimension and $\nu=4/3$ is is the correlation-length exponent, both known exactly in two dimensions. This analysis, however, leaves open the question of the
scaling behavior of $a$ in other regimes, particularly
including the biologically relevant near-ordered behavior
where $\gamma \to \infty$.

In this paper we develop novel theoretical insights about the the
statistical topology of random streamlines in 2D biased random fluid
flow (or magnetic field) for the full range of the control parameter
$\gamma$. We first consider properties that are independent of the
specific ensemble of flow fields, and identify a general upper bound
on the trapped area, based on considerations about the expected value
and variance of the velocity on open and closed streamlines. These
insights are then extended to a more specific result, leading to an
exact expression for the trapped area fraction $a$ in terms of the
variance in the asymptotic displacement of the flow field. We then
derive a formal power series for this quantity in terms of moments of
the flow field. We subsequently derive a tighter bound on the trapped
area for the narrower class of flows described by a Gaussian ensemble
using a different approach specific for this ensemble. Finally, we
provide numerical data for the trapped area in the case of a
particular Gaussian flow ensemble, and verify that the bounds we
established are satisfied. We observe that the second bound accurately
captures the asymptotic statistics of the trapped area in the
large-$\gamma$ regime.

\begin{figure}
    \centering
    \includegraphics[width=.7\linewidth]{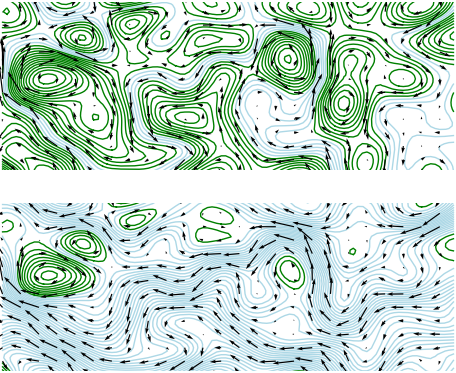}
    \caption{Segments sampled from a random 2D flow field. The
      streamfunction for the flow field was drawn from a Gaussian
      process with zero mean and a radial basis function covariance
      matrix, and a constant mean value was added to the derived
      velocity. Closed streamlines are colored in green, while open
      streamlines are colored in blue. The \emph{upper panel} shows a flow
      with $\gamma = 0.1$, while in the \emph{lower panel} 
      $\gamma = 0.7$.}
      \label{fig:flow-field}
\end{figure}

\section{Results}
\subsection{Analyzing the trapped area}

Consider a zero-mean-flow ensemble from which we draw a
divergence-free 2D vector field $\vect v_1$, and a constant vector
$\mu \hat{\vect v}_0$, with $\hat{v}_0$ taken to be a unit vector in an
arbitrary direction, so that $\mu$ represents the magnitude of the mean flow.
We will add $\vect v_0$ to our zero-mean field $\vect v_1$
to produce a biased random flow field $\vect v$. It will be useful for our
purposes to consider the streamfunction in addition to the underlying flow
field, which is the function $\Phi$ defined by the relationship $\nabla \Phi =
R_{\pi/2} \nabla v$, where $R_{\pi/2}$ is the $\pi/2$-rotation operator in the
plane. The orbits of the flow field are associated with the level sets of this
function, so the problem of studying the orbits can be reduced to the problem
of studying the level sets of $\Phi$. For the biased random flow field
described above, the streamfunction is given by
$\Phi =\Phi_1 + \mu R_{\pi/2} \hat{\vect v}_0 \cdot \vect x$, where $\Phi_1$ is the streamfunction associated with the unbiased portion of the flow. We will define
$\vect \varphi_t(\textbf{x})$ to be the trajectory of a particle at an initial position $\textbf{x}$ advected by the flow for a time $t$, i.e. the map defined such that $\partial_t \varphi_t(\textbf{x}) = \textbf{v}(\varphi_t(\textbf{x}))$. The central question we will
be interested in is the following: how does the percentage of trapped
area $a$ (\emph{i.e.}, the area that does not lie on an open, `escaping'
trajectory) change as the magnitude of the mean velocity $\mu$
increases?

We can leverage a relatively simple insight to obtain an upper bound
on the trapped area. Consider a closed orbit $\partial B$ of the
velocity, which represents the boundary of a trapped region $B$. The
mean velocity in this area is
\begin{equation}
  \int_B \vect v \,\dd^2 \vect x = -R_{\pi/2} \int_B \vect \nabla
  \Phi\,\dd^2 \vect x, 
\end{equation}
$\partial B$ is by definition a closed orbit of the flow field, so the streamfunction $\Phi$ is level on the boundary. The integral of $\nabla \Phi$ over $B$ is thus the integral of a total derivative of a function over a region on whose boundary the function is level, which therefore must be zero.

Since the global mean flow is $\mu\hat{v}_0$, we can derive the mean flow on untrapped areas as well as for trapped areas. Taking $a$ to be the proportion of the plane lying on a closed streamline, we find that the mean flow on open streamlines must be $\mu \hat{v}_0/(1 - a)$. By itself this observation does not allow us to infer anything about $a$. However, the fact that we have two separate regions in the plane with known mean values for the flow field lets us place a lower bound on the variance of the flow field. This variance is the Bernoulli variance associated with being on either the trapped or untrapped area.

We formalize this below. Let $A$ be the set of points on closed streamlines. We can now decompose
the velocity as follows:
\begin{equation}
  \vect v(x) =  \frac{\mu \hat{\vect  v_0}}{1 - a} I[\vect x \notin
  A]+ \vect v(\vect x) I[\vect x \in A] \\ 
\quad + \left(\vect v(\vect x) - \frac{\mu \hat{\vect v_0}}{1 -
    a}\right) I[\vect x \notin A], 
\end{equation}
where  $I[B]$ is the indicator function of set $B$. The first term represents the mean flow on each topologically-defined area, taking the value $0$ on trapped areas and $\mu \hat{v}_0/(1 - a)$ on untrapped areas. The second and third term represent the mean-subtracted value of the flow on a specific topological area.  Each of these terms represents a random variable with respect to the
sampling of $\vect x$, with the first representing a Bernoulli random
variable with the probability $a$ and magnitude
$\mu \hat{\vect v_0}/(1-a)$, and the second two representing zero-mean
random variables. Taking the variance of the velocity magnitude over
space $\sigma^2 = \langle \vect v\cdot \vect v \rangle - \mu^2$, we
get
\begin{equation}
  \sigma^2 = \frac{\mu^2 a}{1 - a} + a\sigma_{A}^2 + (1 - a) \sigma_{\neg A}^2,
\end{equation}
where $\sigma_A^2$ and $\sigma_{\neg A}^2$ represent the conditional
variances of velocity in the trapped and the escaping areas
respectively. The last two terms are strictly positive, yielding the
inequality $\sigma^2 \geq \frac{\mu^2 a}{1 - a}$. Rearranging yields
the following bound:
\begin{equation}
  a \leq \frac{1}{1 + \gamma^2}\,,
\end{equation}
where $\gamma = \mu/\sigma$ is the dimensionless parameter that
quantifies the ratio of `order' to `disorder' of the velocity field.

\subsection{An exact expression for the trapped area}

We can extend the insights above to get an expression for the trapped
area. Let $\vect d_t(x) = \vect \varphi_t(x) - \vect x$ denote the
displacement of a point $\vect x$ under the flow at time $t$. The
conditional average displacement on the trapped and the escaping areas
are $\vect 0$ and $\mu \hat{\vect v}_0 t/(1 - a)$ correspondingly. The
variance of displacement $\langle \vect d_t\cdot \vect d_t\rangle$
must grow slower than $t^2$. Indeed, this variance should be time
reversible (\emph{i.e.}, for large $t$ we should not be able to distinguish
$\vect d_t\cdot \vect d_t - \mu^2 t^2$ from
$\vect d_{-t}\cdot \vect d_{-t} - \mu^2 t^2$), thus it must be either
$\mathcal O(1)$ or $\mathcal O(t^2)$ for $t\to\infty$. If the variance
were asymptotically $\mathcal O(t^2)$, a measurable proportion of
trajectories would diverge by arbitrarily large distances from their
mean, which occurs with probability zero. Thus asymptotically this
variance is $\mathcal O(1)$.

This yields the following asymptotically exact formula for the trapped
area:
\begin{equation}
  \lim_{t \to \infty} \frac{1}{t^2} \langle \vect d_t \cdot\vect  d_t \rangle -
  \mu^2 
  = \frac{\mu^2 a}{1 - a}\,,
\end{equation}
where $\vect d_t$ is the displacement at time $t$ and $\mu$ is the
square of the expected velocity.  Therefore
\begin{equation}
    a = \frac{1}{1 + \mu^2/\kappa}\,,
\end{equation}
where
$\kappa = \lim_{t \to \infty} \langle\vect d_t \cdot\vect d_t
\rangle/t^2 - \mu^2$. Thus, if we can compute $\kappa$, we can compute
the trapped area.

To get a handle on this expression, we can take the Taylor expansion
of the displacement:
\begin{equation}
  \vect  d_t(x) = \sum_{n = 1}^\infty \frac{t^n}{n!} \partial_{t=0}^n \vect d_t(x)
  = \sum_{n = 1}^\infty \frac{t^{n}}{n!} (\vect v \cdot \vect \nabla)^{n-1}\vect  v.
\end{equation}
Subsequently we have
\begin{equation}
  \langle\vect  d_t \cdot \vect d_t \rangle =
  \sum_{n = 1}^\infty \sum_{m = 1}^\infty \frac{t^{n + m}}{n!m!}
  \langle \mathcal{K}_{\vect v}^{n-1} \vect v \cdot \mathcal{K}_{\vect v}^{m-1}\vect  v \rangle,
\end{equation}
where $\mathcal{K}_{\vect v} := \vect v \cdot \vect \nabla$. The objects
$\langle \mathcal{K}_{\vect v}^{n-1} \vect v \cdot \mathcal{K}_{\vect
  v}^{m-1} \vect v \rangle$ are of fundamental interest for
determining the trapped area. Note that
$\mathcal{K}_{\vect v}^\dagger = -\mathcal{K}_{\vect v}$ (the flow
operator is unitary as it preserves the Lebesgue measure; therefore,
its generator $\mathcal{K}_{\vect v}$ must be anti-Hermitian). We
have, therefore,
\begin{equation}
  \langle \mathcal{K}_{\vect v}^{n-1} \vect v \cdot \mathcal{K}_{\vect v}^{m-1} \vect v \rangle
  = (-1)^{n-1} \langle \vect v, \mathcal{K}^{n+m-2}_{\vect v} \vect v \rangle.
\end{equation}
Hence,
\begin{eqnarray}\label{eq:series}
  \langle \vect d_t \cdot \vect d_t \rangle &=& \sum_{n = 2}^\infty \frac{t^n}{n!} \langle\vect  v, \mathcal{K}^{n-2}_{\vect v} \vect v \rangle  \sum_{m = 1}^{n-1} (-1)^{m-1} {n \choose m} \nonumber\\
    &=& \sum_{n = 0}^\infty \frac{2t^{2n+2}}{(2n+2)!} \langle \vect v, \mathcal{K}_{\vect v}^{2n}\vect  v \rangle.
\end{eqnarray}
Note that when we take the time-average of
$\langle\vect d_t \cdot \vect d_t \rangle/t^2$, the flow of time
becomes equivalent to a rescaling transformation, since
$\mathcal{K}_{\vect v}$ is a covariant operator, and the coefficient
in front of the term involving $\mathcal{K}^{2n}$ is $t^{2n}$.

Evaluating the series above for a generic flow ensemble is
difficult. One flavor of this difficulty is discussed in~\ref{appendix:series},
where we present an initial
sketch of the calculations required for a particular Gaussian
ensemble. Evaluating each term often requires a combinatorial
expansion that becomes intractably difficult as the order of the terms
increases; and, even in cases where simplifications arise, we have no
guarantee that the resultant expressions will lend themselves to the
kind of asymptotic analysis required to derive the trapped area.

\subsection{An upper bound on the trapped area for the Gaussian ensemble}

While we have an upper bound on the trapped area in the form
$a \leq 1/(1 + \gamma^2)$ from simple considerations about the
variance of the velocity, we can obtain an improved bound by using a
different approach in the case where the flow field is drawn from an ensemble
with known secondary characteristics. Here, we will consider the case of a flow
field $\textbf{v}$ wherein the marginal distribution for each component of the
velocity, conditioned on a single point in space, is a uniform Gaussian. We
will assume that the $\hat{v}_0$-parallel component of the field has the
marginal distribution $\mathcal{N}(\mu, \sigma/\sqrt{2})$, and the
$\hat{v}_0$-orthogonal component the marginal distribution $\mathcal{N}(0,
\sigma/\sqrt{2})$. This scaling ensures consistency with our previously given
definitions of $\mu$ and $\sigma$ and preserves the rotational invariance of
the unbiased component. 
This new bound leverages the simple insight that
$\langle\vect v(\vect x) \rangle_{ \vect x \in A} = 0$. Since $A$ is a
set with zero expected velocity, its area fraction $a$ must be no larger
than the area fraction of the maximal set $S$ such that
$\langle\vect v(\vect x) \rangle_{\vect x \in S} = 0$. We will define
this area fraction as $s$, and write our bound as $s \geq a$.
\begin{figure}
    \centering
    \includegraphics[width=\linewidth]{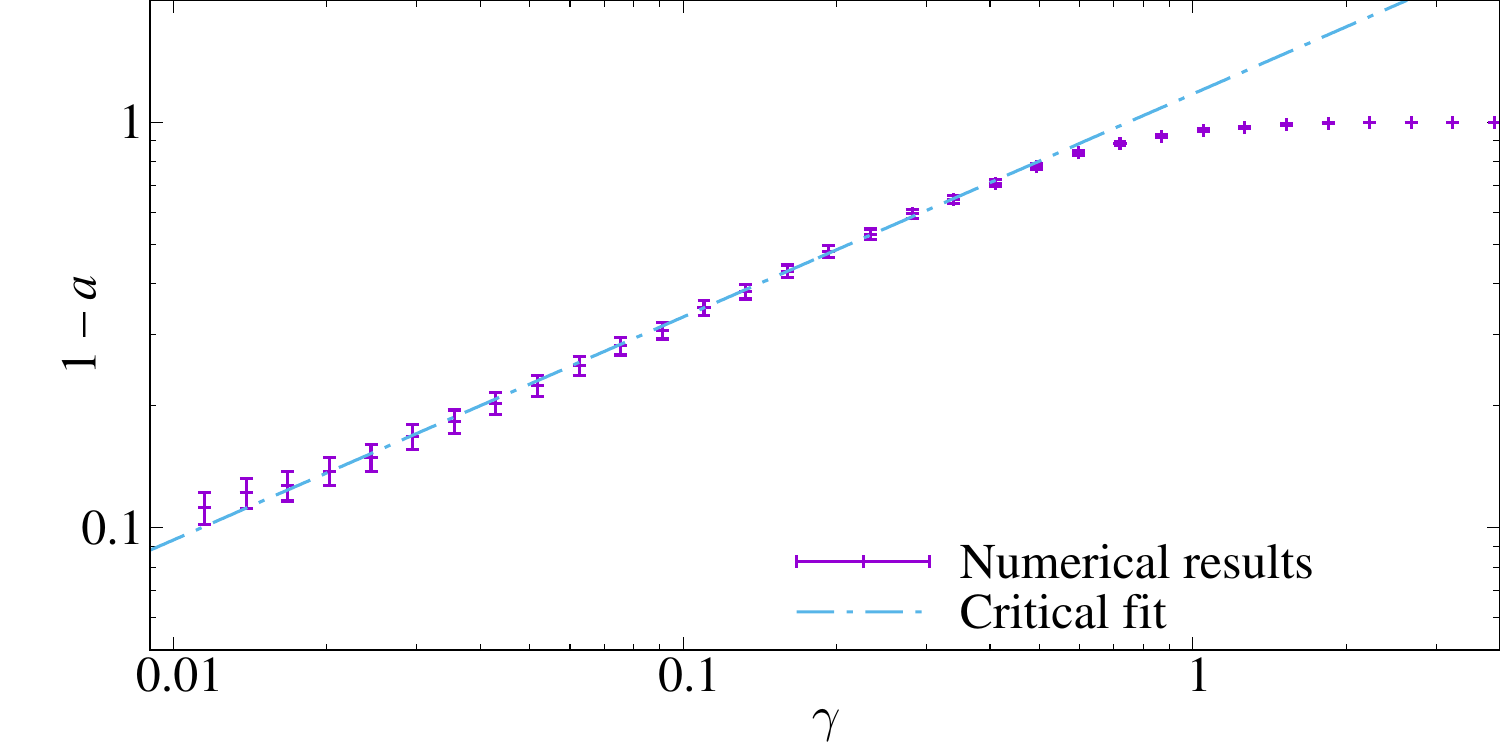}
    \caption{Numerical results for the trapped area in the critical regime  for the Gaussian flow ensemble. The straight line is a fit to $1-a=B \gamma^c$ for $0.01\leq\gamma\leq0.3$, where $B$ is a constant and $c=0.552(13)$, compatible with the value $c=4/7$~\cite{isichenko1992percolation}. The goodness-of-fit $\chi^2$ parameter per degree of freedom is
    $\chi^2 /\mathrm{d.o.f.}=3.96/16$. For $\gamma>2$ we do not find any open streamlines in our 1000 samples.
    \label{fig:critical}}
\end{figure}
\begin{figure}
    \centering
    \includegraphics[width=\linewidth]{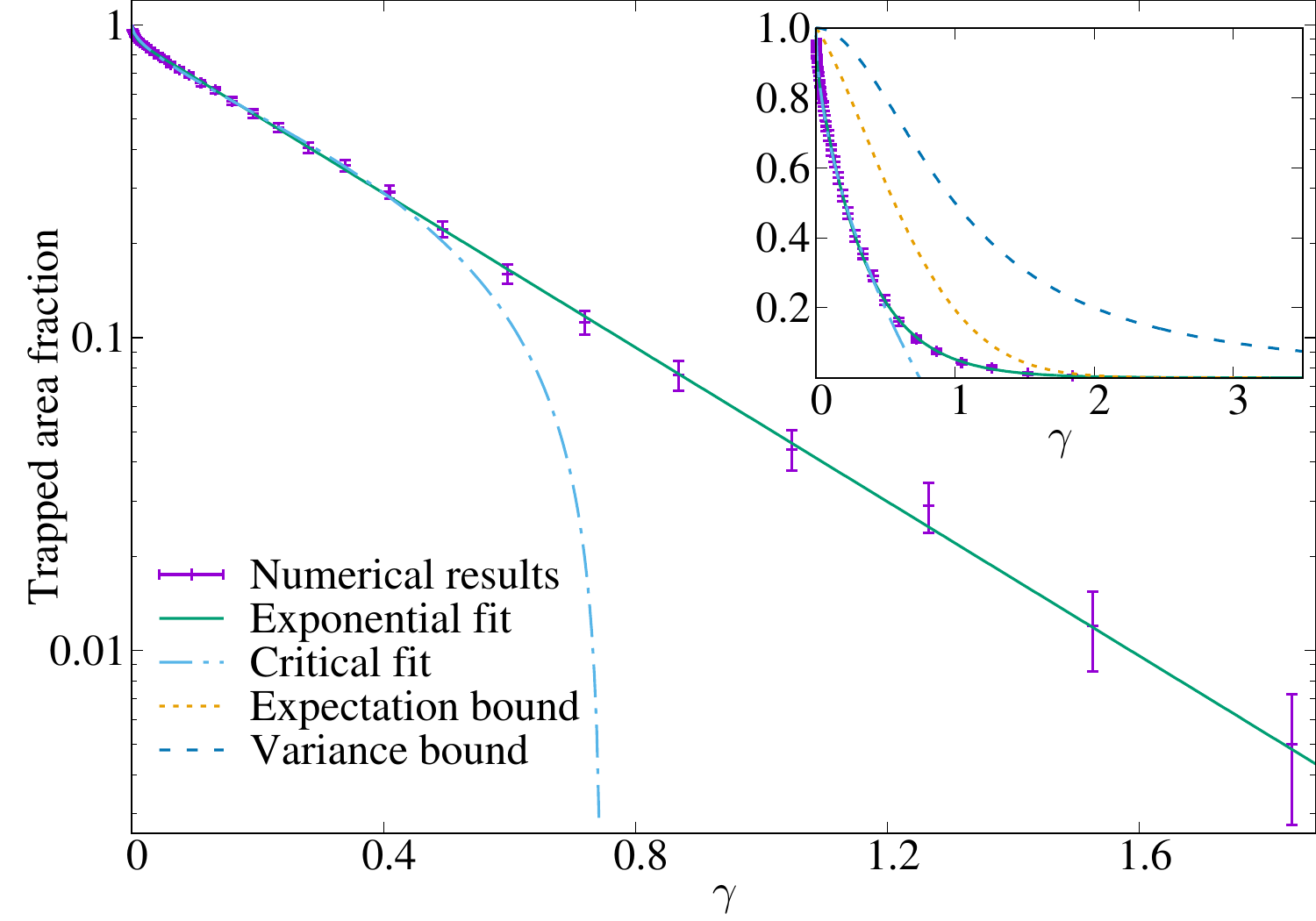}
    \caption{Numerical results for the trapped area in the ordered and disordered regimes. The \emph{inset} shows our
    full range of simulations in a linear scale, compared with the two general analytical bounds computed in this paper, with the critical fit of Fig.~\ref{fig:critical} and with an exponential fit. The curve crosses over from the critical behavior of Fig.~\ref{fig:critical} when $\gamma\to0$ to the $a\sim \ee^{-\gamma^2}$ behavior of the expectation bound when $\gamma\to\infty$. In the intermediate regime, exponential behavior is observed.
    \label{fig:trapped-figure}}
\end{figure}

Note that, for a velocity component $v_x \sim p$ with a
positive expected value, the largest subset we can achieve with zero
expected velocity is given by selecting all values of $v_x \leq k$ for
some $k$ such that $\int_{-\infty}^k v_x p(v_x)\,\dd v_x = 0$. The
fraction of the total area belonging to this subset is then given by
the value of the CDF at $k$, \emph{i.e.} $\int_{-\infty}^k p(v_x)\, \dd
v_x$. In our case, the component of the velocity aligned with
$\hat{v}_0$ is distributed according to
$v \sim \mathcal{N}(\mu, \sigma/\sqrt{2})$, so we have the conditions
\begin{equation}
  0 = \int_{-\infty}^k v \exp\biggl(-\frac{(v - \mu)^2}{\sigma^2}\biggr)\,\dd v 
  = \sqrt{\pi} \mu \, \big(\mathrm{erf}(\frac{k - \mu}{\sigma}) +
  1\big) - \sigma \ee^{-(k-\mu)^2/\sigma^2},
\end{equation}
In terms of $k$ we can write $s$ as
\begin{equation}
  s = \frac{1}{2} \biggl[1 + \mathrm{erf}\biggl(\frac{k - \mu}{\sigma}\biggr)\biggr]\,.
\end{equation}
Solving these equations numerically for $s$ gives us an upper bound on
$a$ for the Gaussian ensembles. In~\ref{appendix:expectation}
we show that this bound goes as $s \sim \mathcal{O}(\gamma^{-1}e^{-\gamma^2})$ when $\gamma\to \infty$. If we find ourselves in possession of a convenient expression for the CDF of another ensemble, we can apply similar reasoning derive an upper
bound for the trapped area in that ensemble.

This analysis also reveals a procedure for generating progressively
tighter bounds on $a$. $A$ is defined to be precisely the set where
the expected displacement is zero for all values of $t$. This entails
that each derivative
$\partial_n^t \vect d_t(x) = [\vect v\cdot\vect \nabla]^n \vect v$ is
zero. If we can identify the marginal distribution for each
derivative or, better, the joint distribution between all
derivatives, we can follow a similar argument to identify the
largest set that has zero expected value for each derivative
considered.

\subsection{Numerical results}
In order to investigate the trapped area numerically, we start by generating
values of $\gamma$ uniformly spaced
in a logarithmic scale in the range $10^{-2}\leq \gamma \leq 4$.
Subsequently, for each $\gamma$, we initialize 1000 random zero-mean streamfunctions, using a procedure that we will describe, compute the value of $\mu$ associated with their variance, and generate a velocity field by combining the streamfunction gradient and the mean velocity. For each streamfunction we then integrate the trajectory that the point $\vect{x}_0 = (0,0)$ (the statistical ensemble from which the velocity fields are drawn is translationally invariant, so the initial point on the trajectory can always be taken to be the origin without loss of generality) follows under this velocity field over a time range of $t = 600$, using the function odeint in the Python package scipy. We then determine whether each trajectory is open or closed by examining whether the closest distance it approaches to the origin at large times exceeds a critical threshold, in which case we classify it as an open trajectory, and otherwise consider it closed. The proportion of closed trajectories out of the 1000 samples becomes our estimate for the trapped area for each value of $\gamma$. The choice of a very large time range of $t = 600$ ensures that there is no ambiguity between open and closed trajectories, as it is much larger than the period of all closed curves in our samples.

In order to generate our streamfunctions, we note that, in the infinite-size limit, a
stationary Gaussian process can be characterized exactly by its power
spectral density. That process can, therefore, be sampled by
choosing a random phase for each wavevector. We employ
unbiased streamfunction of the form
\begin{equation}
\Phi_1(\vect x) = \sum_{|\vect k| \leq 150} \exp\biggl(-\frac{\vect k^2}{2\lambda^2}\biggr) \cos\bigg(\frac{2\pi}{L} \vect k \cdot \vect x - \phi_k\bigg),
\end{equation}
where $\lambda$ is a correlation-length parameter and $L$ is a parameter that picks the maximum wavelength that enters into the streamfunction. This approach approximates sampling a streamfunction with a Gaussian radial basis function covariance kernel in the large-size limit. Each phase offset $\phi_k$ is sampled uniformly from the interval
$[0,2\pi]$. In this work, we choose $L=5$ and $\lambda=10$. This approach to sampling a streamfunction, as opposed to directly sampling a Gaussian process on a lattice, is beneficial because  a) it is computationally efficient relative to direct sampling, which for an $N \times N$ lattice requires computing the Cholesky decomposition of a covariance matrix of size $N^2\times N^2$ [an $\mathcal O(N^6)$ operation]; b) it avoids discretization issues that could occur when integrating a trajectory derived from a velocity field sampled on a lattice; and c) it can be naturally extended to an arbitrary distance from the origin, which allows us to ignore boundary issues at the lattice edge. This approach, however, suffers from the potential drawback that it does not fully capture the statistics of the Gaussian process outside of the bandwidth used for sampling, which means that it is inapplicable for very small $\gamma$ (when the average loop size becomes smaller than the minimum wavelength) and for very large $\gamma$ (when the average loop size exceeds the finite system size). In practice, see below, we find that a large range of $a$ can be accessed for $\gamma$ that avoid these limits.

We first examine the scaling of the trapped area in the critical ($\gamma\to0$) regime in Fig.~\ref{fig:critical}, which is expected to follow the law
$1-a \propto \gamma^c$, with $c=4/7$~\cite{isichenko1992percolation}. In the $\gamma\leq 0.3$ range, we find a best-fit exponent of $c=0.553(13)$, compatible with the value $c=4/7 \approx 0.571\ldots$ For larger $\gamma$, however, the trapped area strongly deviates from this behavior. Indeed, as discussed above and in~\ref{appendix:expectation}, in the $\gamma\to\infty$ limit, the trapped area is bounded by $a \leq \ee^{-\gamma^2}/{(2\gamma\sqrt{\pi})}$. Between these two limiting behaviors, we find empirically that there is a very robust exponential dependence of the form $a \approx \exp(-d\gamma)$, with a best-fit value of $d=2.84(4)$, which we illustrate in Figure~\ref{fig:trapped-figure}~\footnote{We actually consider a sum of two expontials, $a=A_1 \ee^{-d_1 \gamma} + A_2 \ee^{d_2 \gamma}$. This effective function fits all of our data up to $\gamma \approx 2$.}. While there are no evident analytical simplifications that would allow us to predict this behavior directly, we note that intuitively this scaling means that the rate of decrease of a given trapped region as $\gamma$ increases is roughly proportional to the area contained inside. In  short, the trapped area is effectively described
by an exponential dependence on $\gamma$ for a wide range.

Figure~\ref{fig:trapped-figure} also shows our analytical bounds on the trapped area. The first important thing to notice is that the `variance bound' $a \leq 1/(1 + \gamma^2)$ is conservative. In our computations the trapped area decreases at a significantly faster rate. The bound offered by considering the largest set with zero expected velocity (the `expectation bound') is much tighter.

\section{Discussion}
In this paper we have discussed the problem of determining the fraction of
the `trapped area', \emph{i.e.}, the area of topologically closed
trajectories, for random two-dimensional flow fields. Our analysis extends beyond
the (critical) disordered limit, which can be understood using the percolation-theory 
universality class, and into the ordered regime which, moreover, is the more
relevant for many applications. We introduce
several techniques that lead both to a general upper bound
for a broad class of flow ensembles and to a specific upper bound
for Gaussian flow fields. We also derive a power-series expression
for the trapped area using asymptotic analysis of the displacement
variance under the flow map. Finally, we compute the trapped area 
numerically for a Gaussian flow field and find that beyond the critical regime
it is well approximated by an exponential.

In addition to providing theoretical insights for the features of
interest in two-dimensional flows, this work has a number of broader
implications. First, we expect that the techniques introduced in this
paper could be extended to allow an estimation or even an exact
computation of other topological features of fluid flows, potentially
even in higher-dimensional regimes. Second, in some applications of
fluid flows and their close cousins, magnetic fields, the topology of
orbits plays a crucial role in the overall behavior of the system. The
orbit topology in magnetic materials has been related to certain
electrical properties and to the Hall effect
\cite{trugman1983localization,isichenko1992percolation}. Orbit
topology has implications for the behavior of ciliated flows in
biological systems, ranging from mucosal transport in the airway to
the feeding of starfish larvae \cite{RamirezSanJuan2020,
  ding2015selective, gilpin2020multiscale,
  GilpinVivekPrakash2017}. Being able to compute the
proportion of area trapped in closed streamlines will enable
analytical investigations into the fundamental principles and
tradeoffs underling these systems. In particular, this analysis
may have significant implications for understanding
the progress of infections in the airway, where areas of trapped
flow provide a foothold for viral infections
\cite{RamirezSanJuan2020, cilia2021}.

\paragraph*{Data availability} The code and data used to 
generate the figures in this paper can be accessed at:\newline
\url{https://github.com/Kambm/TrappedAreas}.

\section*{Acknowledgments}
The authors acknowledge the generous support of Chan Zuckerberg
Initiative and Chan Zuckerberg Biohub. D.Y.
acknowledges support by MINECO (Spain) through Grant No. PGC2018-094684-B-C21,
partially funded by the European Regional Development Fund (FEDER).

\appendix
\section{Analyzing the series}\label{appendix:series}
Here we present a sketch of the calculations required to evaluate the series \ref{eq:series} exactly. We do not complete this here but instead provide a partial solution in order to give the reader a sense of the difficulties involved.

We can express the term $\langle v, \mathcal{K}^n_v v \rangle$ using
Einstein notation:
\begin{equation}
  \langle v, \mathcal{K}_v^{2n} v \rangle = v_i \cdots v^l \partial_l [v^k \partial_k [v^j [\partial_j v^i]]].
\end{equation}
To evaluate these terms we need to sum over all ways to distribute the
derivatives across each term. We can illustrate this symbolically as
follows:
\begin{equation}
    \left|
    \begin{array}{ccccc}
    \cdots & v^l & v^k & v^j & v^i\\
    \hline
    & & & & \partial_j\\
    & & & \partial_k & \partial_k\\
    & & \partial_l & \partial_l & \partial_l\\
    & \partial_m & \partial_m & \partial_m & \partial_m\\
    \reflectbox{$\ddots$} & & & & \vdots
    \end{array}
\right|
\end{equation}
We then sum over each allowed choice of placement for each derivative
term. We also note that $v = R_{\pi/2} \nabla \Phi$ (or equivalently
$v^i = (-1)^i \partial_{1 - i} \Phi$), so we can ultimately express
the entire term using derivatives of the $n$-point correlation function
for $\Phi$ at the origin. \emph{E.g.}:

\begin{equation}\fl
  \langle v_i v^k \partial_k v^j \partial_j v^i \rangle
  = (-1)^{j + k}\frac{\partial}{\partial w_{1-i}} \frac{\partial}{\partial z_{1-k}} \frac{\partial}{\partial y_k} \frac{\partial}{\partial y_{1-j}} \frac{\partial}{\partial x_j} \frac{\partial}{\partial x_{1-i}} C^4(w,z,y,x) \bigg|_{x,y,z,w = 0}\,,
\end{equation}
where
\begin{equation}
  C^n(w,x,y,z,\cdots) = \langle \underbrace{\Phi(w)\Phi(x)\Phi(y)\Phi(z)\cdots}_{n\ \mathrm{times}}\rangle.
\end{equation}
In order to proceed further we will need to get a handle on the
correlation function. When $\mu = 0$, we can employ Wick's theorem to
express the even-order correlation functions as a sum over products of
second-order correlations:
\begin{equation}
  C^{2n} = \prod_{i = 1}^n \langle \Phi(x_{2i - 1})\Phi(x_{2i}) \rangle + \mathrm{other\ pairings}
\end{equation}
When $\mu \neq 0$, we can can write
$C^{n}= \langle (\Phi_1(x) + \mu R_{\pi/2}\hat{v}_0\cdot x) (\Phi_1(y)
+ \mu R_{\pi/2}\hat{v}_0\cdot y) \cdots \rangle$. Distributing terms
gives
\begin{equation}
  C^{2n} = \sum_{S \subset V} C_0^{2n - |S|}(V - S) \prod_{x \in S} (\mu R_{\pi/2}\hat{v}_0 \cdot x)
\end{equation}
where $V = \{x_1, \cdots, x_{2n}\}$ and $C_0^n$ refers to the $n$-point
correlation function for the unbiased streamfunction. To proceed
further, we will have to posit a form for the unbiased two-point
correlation function. We will choose the form
\begin{equation}
  C(x,y) =
  \frac{\lambda^2\sigma^2}{2}\exp\biggl(-\frac{\Vert{x-y}\Vert^2}{2\lambda^2}\biggr),
\end{equation}
where $\lambda$ is the characteristic length. This choice ensures that
$\langle v\cdot v \rangle = \mathrm{Tr}[\nabla_x \nabla_y C(0,0)] =
\sigma^2$. What happens when we differentiate this at the origin? The
$2k$th derivative of the Gaussian $\exp(-\lambda^2x^2/2)$ at the
origin is $(2k)!(-1)^k/\bigl((2\lambda^2)^kk!\bigr)$, while all odd
derivatives are zero.

Now we consider evaluating a term with a specific arrangement of
auxiliary derivatives. We can categorize each such term using a list
of tuples $(a_i, b_i)$ for $i = 1, \cdots, 2n+2$, where the tuple
$(a_i,b_i)$ indicates that the operator
\begin{displaymath}
  \frac{\partial^{a_i}}{\partial x_0^{a_i}}
  \frac{\partial^{b_i}}{\partial x_1^{b_i}}
\end{displaymath}
is applied to the $i$th argument. We then pair each argument with
another, apply the respective derivatives to the two-point correlation
function associated with the paired arguments, take the product of all
the results, and sum over all possible pairings. The contribution from
the isotropic component of the two-point correlation function, from
each pair $(i,j)$, is
\begin{eqnarray}
\fl
  (-1)^{(a_i - a_j) + (b_i - b_j)} \frac{\lambda^2 \sigma^2}{2} \frac{2(a_i + a_j)!}{(a_i + a_j)!} \frac{2(b_i + b_j)!}{(b_i + b_j)!} \frac{(-1)^{a_i + a_j + b_i + b_j}}{(2\lambda^2)^{a_i + a_j + b_i + b_j}}
  =\nonumber \\ \qquad\qquad\qquad\qquad\frac{\lambda^2 \sigma^2}{2} \frac{2(a_i + a_j)!}{(a_i + a_j)!} \frac{2(b_i + b_j)!}{(b_i + b_j)!} \frac{1}{(2\lambda^2)^{a_i + a_j + b_i + b_j}}\,.
\label{eq:isotropic-component}
\end{eqnarray}
Eq.~(\ref{eq:isotropic-component}) assumes that $a_i + a_j$ and $b_i + b_j$ are even; otherwise the
result is 0.  We see that the parity associated with the overall order
of the derivatives and the parity associated with the pairing of
arguments for each component cancel out. If the pairing is specifically
(0,1), (0,1), the result from the deterministic component is $\mu^2$
(we assume without loss of generality that $\hat{v}_0 = \hat{x}$), and
otherwise the deterministic component contributes nothing.

In order to complete the analysis, one would like to solve the combinatorial problem of summing these coefficients over derivative pairings. This is ultimately a difficult problem, and the authors leave it to the interested reader to pursue it in further depth.  

\section{Asymptotic scaling of the expectation bound}\label{appendix:expectation}

The equations that determine the expectation bound are
\begin{eqnarray}
    0 &=& \, \sqrt{\pi}\biggl[\mathrm{erf}(\frac{k - \mu}{\sigma}) + 1\biggr] - \gamma^{-1}
      \ee^{-(k-\mu)^2/\sigma^2},\\
    s &=& \frac{1}{2} \biggl[1 + \mathrm{erf}\biggl(\frac{k - \mu}{\sigma}\biggr)\biggr]\,.
\end{eqnarray}
We will define $Q = \frac{\mu - k}{\sigma}$ to be the negative argument in the various exponentials. Rewriting the first equation gives
\begin{equation}
    \gamma = \frac{e^{-Q^2}}{\sqrt{\pi}(\mathrm{erf}(-Q) + 1)}.
\end{equation}
For large $Q$, $\mathrm{erf}(-Q) \sim \frac{e^{-Q^2}}{Q\sqrt{\pi}}-1$, so this expression reduces to
\begin{equation}
    \gamma \sim \frac{e^{-Q^2}}{\sqrt{\pi}\frac{e^{-Q^2}}{\sqrt{\pi}Q}} = Q.
\end{equation}
Substituting this into our relation for $s$ gives the asymptotic behavior of the bound:
\begin{equation}
    s \sim \frac{e^{-\gamma^2}}{2\gamma\sqrt{\pi}}.
\end{equation}

\section*{References}
\bibliographystyle{iopart-num}

\providecommand{\newblock}{}

\end{document}